\documentstyle[preprint,aps]{revtex}
\let\eps=\epsilon \let\a=\alpha 
\let\s=\sigma \let\ph=\varphi \let\R=\right \let\L=\left \let\q=\quad
\let\qq=\qquad \let\e=\varepsilon \def\cD{{\cal D}} \def\cS{{\cal S}}
\def\cZ{{\cal Z}} \newcommand{\ts}{\textstyle}
\newcommand{\rf}[1]{(\ref{#1})} \newcommand{\lb}[1]{\label{#1}}
\renewcommand{\v}[1]{{\vec{#1}}}
\def\be{\begin{equation}} \def\ee{\end{equation}}
\def\bea{\begin{eqnarray}} \def\eea{\end{eqnarray}}
\def\ba{\begin{array}} \def\ea{\end{array}}

\begin{document}
\renewcommand{\thefootnote}{\fnsymbol{footnote}}
\setcounter{footnote}{1}
\title{ The Heisenberg
antiferromagnet on a triangular lattice:\\ topological excitations}

\author{M.\ Wintel and H.\ U.\ Everts}
\address{Institut f\"ur Theoretische Physik, Universit\"at Hannover\\
Appelstr.\ 2, D-30167 Hannover, Germany}

\author{W.\ Apel}
\address{Physikalisch-Technische Bundesanstalt\\
Bundesallee 100, D-38116 Braunschweig, Germany}
\date{\today}

\maketitle
\begin{abstract}
We study the topological defects in the classical Heisenberg antiferromagnet
in two dimensions on a triangular lattice (HAFT).
While the topological analysis of the order parameter space indicates that the
defects are of $Z_2$ type, consideration of the energy leads us to a
description of the low--energy stationary points of the action in terms
of $\pm$ vortices, as in the planar XY model.
Starting with the continuum description of the HAFT, we show analytically
that its partition function can be reduced to that of a 2--dimensional
Coulomb gas with logarithmic interaction.
Thus, at low temperatures, the correlation length is determined by the
spinwaves, while at higher temperatures we expect a crossover to a
Kosterlitz--Thouless type behaviour.
The results of recent Monte Carlo calculations of the correlation length
are consistent with such a crossover.
\end{abstract}
\vspace{5mm}


The effect of frustration on the properties of two--dimensional
antiferromagnets has been the subject of numerous investigations [1--7].
The Heisenberg antiferromagnet on a triangular lattice (HAFT) is a
prototypical model for such systems.
In early studies \cite{and1}, the focus has been on the ground state properties
of the quantum mechanical model.
These properties are qualitatively similar to those of the same model on a
square lattice (HAFSQ) \cite{cha,man} where there is no frustration.
What will be of interest here and possibly also for the interpretation of
experimental results is the behaviour at finite temperatures.
As is frequently observed, the HAFT possesses vortex excitations,
which are important at finite temperatures and which are absent in the case of
the HAFSQ.
The reason is the difference in the order parameters.
On the triangular lattice, in contrast to the square lattice case,
the order of the spins at zero temperature is not collinear;
instead, the frustration leads to a state in which, on every elementary
triangular plaquette, the spins form a regular planar star.
As a consequence, the order parameter space is isomorphic to the full
3--dimensional rotation group $SO(3)$ (HAFSQ: $SO(3)/SO(2)=S_2$).
The $SO(3)$  (in contrast to the $S_2$)  is not simply connected, and a
homotopy analysis predicts the existence of topological stable point defects.
Because the fundamental homotopy group is $\Pi_1(SO(3))=Z_2$ \cite{mer}
these defects are called $Z_2$--vortices.
The earliest hint as to the importance of the vortices came from the
Monte Carlo simulations of the classical HAFT by Kawamura and Miyashita
\cite{km}.
They found evidence for a defect--driven transition which is strongly
reminiscent of the Kosterlitz--Thouless transition
\cite{kt} in the planar XY model.
However, there, in the $SO(2)$ case, the vortices are characterized by
integer numbers ($\Pi_1(SO(2))=Z$) \cite{mer}.
{}From their data, Kawamura and Miyashita cannot conclude whether the
transition is of Kosterlitz--Thouless type or if it is caused by
a different mechanism.

The purpose of this letter is to treat analytically the vortices in the HAFT.
It will be shown that, after an integration of the spinwaves in harmonic
approximation, the partition function reduces to that of a 2-dimensional
Coulomb gas with logarithmic interaction.
Thus, we arrive at a similar description as in the case of the XY model.
In the derivation of this result we have to face two problems.
(i) How can one understand the picture of a gas of $\pm$ charged particles,
when one knows from homotopy considerations \cite{mer} that there is only
one type of vortices ?
(ii) In constrast to the case of XY model, the spinwaves in the HAFT
are coupled to the vortices, already in the harmonic approximation.
While the vortices interact logarithmically in the
absence of this coupling, it is not at all trivial to show that this
logarithmic form survives despite of the coupling.
Furthermore, anharmonic spinwave interactions are known to yield a finite
correlation length $\xi_{SW}$ for the HAFT for arbitrarily low temperatures
\cite{awe,aza1}, whereas the correlation length of the XY model
is infinite below the Kosterlitz--Thouless temperature.
As we shall discuss below, this restricts the validity of the logarithmic
vortex interaction to distances $r\ll\xi_{SW}$.

Analogous to the investigation of the planar XY model, our
starting point is the action of the continuum version of the
classical HAFT, i.~e.~the nl$\s$ model \cite{awe,diss}
\be
\cS \{ \v{n}_k \} = - \frac{1}{4t} \int d^2x \L\{
 \L(\v{\nabla}\v{n}_1 \R)^2  + \L(\v{\nabla}\v{n}_2 \R)^2 \R\} \lb{0}
\q,\qq t = {\ts \frac{2}{\sqrt{3}} \frac{k_B T}{J S^2}} \q.
\ee
The orthonormal vectors $\v{n}_1(\v{x})$ and $\v{n}_2(\v{x})$
span the plane of the original planar star of spins on each
elementary triangular plaquette.
Since the $SU(2)$, as the universal covering group
of the $SO(3)$, plays an essential role in homotopy analysis \cite{mer},
we represent the field configuration
by $SU(2)$ fields $g(\v{x})$:
\be
\v{n}_k = \frac{1}{2}\, tr \L\{ \v{\s}\,g\,\s^k g^{-1} \R\}\q , \qq
g(\v{x})\, \eps \, SU(2)\q.\lb{2}
\ee
Here $\s^k\, (k=1,2,3)$ are the Pauli matrices and
$\v{n}_3 = \v{n}_1\! \times \! \v{n}_2$.
In general, the $SU(2)$ fields can be written as
$g(\v{x}) = \exp \L\{ \frac{i}{2}\; \v{u}(\v{x})\!\cdot \! \v{\s}\;\;
\Psi(\v{x}) \R\}$ where $\v{u}(\v{x})$ is a
unit vector specifying a local axis of rotation and $\Psi(\v{x})$ is a local
rotation angle. From \rf{2}, one can see how a configuration
$\v{n}_k (\v{x})\,$ is generated by local rotations $g(\v{x})$.
The ambiguity in the sign of the $SU(2)$ fields in \rf{2} reflects the
existence of two classes of states of different topology.

Next, we determine the stationary points of \rf{0}.
The factorisation of the fields
$g(\v{x}) = g_s(\v{x}) \,\exp\{i\, \v{\e}(\v{x})\cdot\v{\s}\}$
into a saddle point and a spinwave part leads to Euler--Lagrange equations.
The trivial solution $g_s(\v{x})=$\,const \,\, reproduces the spinwave minimum.
Searching for nontrivial solutions $g_s(\v{x})$, we assume
$\v{u}(\v{x})= \v{u}_s = $ const:
$g_s(\v{x}) = \exp \L\{ \frac{i}{2}\; \v{u}_s\!\cdot \!\v{\s}\;\;
\Psi_s(\v{x}) \R\}$.
Minimal action solutions are found to be those which satisfy
$\v{u}_s \cdot \hat{e}_z = 0$.
This ansatz reduces the Euler--Lagrange equations to the Laplace equation
for the field $\Psi_s (\v{x})$.
Thus, $\Psi_s (\v{x})$ must be identical to the vortex
configurations of the $SO(2)$ model. The one--vortex solution centered
at the origin is: $\Psi_1 (\v{x}) = \arctan \frac{y}{x}\,$;
$\;g_1(\v{x})= \exp \{\frac{i}{2}\; \v{u}_s\!\cdot\v{\s}\;\arctan
\frac{y}{x}\}$.
We have proven local stability against small fluctuations $\v{\e}(\v{x})$
explicitly. The energy of one vortex diverges logarithmically with the
system size.

{}From the topological point of view, a $Z_2$ vortex can be characterized by
the
behaviour of the field $g(\v{x})$ on an arbitrary path encircling the
singularity: as one moves around such a closed path, $g(\v{x})$ picks up a
phase factor $e^{i\pi}$, still the fields  $\v{n}_k$  are continuous
corresponding to the above mentioned ambiguity in the sign of the $SU(2)$
fields in \rf{2}.
A general one--vortex configuration is obtained by a continuous deformation of
the stationary configuration $g_1(\v{x})$.
However, in contrast to the $SO(2)$ case, the stationary point $g_1$ in the
homotopy class with one vortex is degenerate with respect to the direction
of $\v{u}_s$ .
Solutions with different orientations $\v{u}_s$ can be continuously transformed
into each other.
In the $SO(2)$ case, vortices and antivortices are stationary points of the
action in different homotopy classes.

Since the Laplace equation for $\Psi_s (\v{x})$ is linear, stationary
N--vortex configurations are obtained simply by linear combination:
\be
 \Psi_N(\v{x}) \;=\; \sum\limits_{j=1}^{N} \;s_j \;
             \arctan \frac{y-y_j}{x-x_j} \q,\qq s_j=\pm 1\lb{7}
\ee
with $N$ singularities at $\v{x}_1,...,\v{x}_N$ and
$g_N(\v{x}) = \exp \L\{ \frac{i}{2}\; \v{u}_s\! \cdot \! \v{\s}\;\Psi_N(\v{x})
\R\}$.
Finiteness of the action demands $\sum_{j=1}^N s_j =0$, i.~e.~$N$ has to be
even.
The value of the action at the stationary point is:
\be
\cS^{(0)}_N \;=\; {\ts \frac{\pi}{2t}}\;\sum\limits_{j,j'} \;s_j s_{j'}\;
             \ln \frac{|\v{x}_j-\v{x}_{j'}|}{a} \q,
\qq (|\v{x}_j-\v{x}_{j'}| \gg a) \lb{7a}\q.
\ee
The vortices interact logarithmically ($a$ is the lattice spacing).
As in the one--vortex case, the $N$--vortex stationary point is degenerate
with respect to the direction $\v{u}_s$.
For any fixed $\v{u}_s$, such a configuration consists of an equal number
of vortices (rotation axes $\v{u}_s$) and antivortices (rotation axes
$-\v{u}_s$).
A more general configuration with an even number of vortices, $g_N$, is
obtained by forming a product of $N$ single vortex fields
$ g_1(\v{x}- \v{x_j}) = \exp \{\frac{i}{2}\; \v{u}_{sj}\!\cdot\v{\s}\;\arctan
\frac{y-y_j}{x-x_j}\}\; (j=1,...,N)$ with independent directions $\v{u}_{sj}$
of the individual vortices.
However, in general, this is not a stationary point of \rf{0}.
We have checked for small $N\,(\leq4)$ that for arbitrary directions
$\v{u}_{sj}$ the energy of such configurations is always higher than
\rf{7a}. Thus in addition to homotopy theory, one needs to consider the energy,
which then singles out the configurations of the form of $g_N$ with equal
number of vortices and antivortices of a fixed axis $\v{u}_s$.
This is the solution to the first problem posed in the introduction.

We now approximate the partition function of our model by summing over the
stationary points and including small fluctuations up to second order:
\be
\cZ \;\approx \;\sum\limits_{N=0}^{\infty}\;
\frac{z^{\frac{N}{2}}}{\L({\ts \frac{N}{2}}!\R)^2}
\;\int d^2 x_1 \cdots \int d^2 x_{N} \;\cZ_{N}\L(\v{x}_1,\cdots,\v{x}_{N} \R)
\lb{8}
\ee
\be
\cZ_{N}\L(\v{x}_1,\cdots,\v{x}_{N} \R)\;:=\;
 \exp \cS^{(0)}_N \,\, \int\! \cD \L[\v{\e}\,\R]\;
 \exp \biggl\{\cS^{(2)}_N \L[\v{\e}\,(\v{x})\R]\biggr\}\q,
\ee
where $z$ is the fugacity of a pair of vortices.
The fluctuation part of the action reads
\be
\cS^{(2)}_N \;=\; -\frac{1}{t}\int\! d^2 x\;
\v{\e}\,(\v{x})^T \;M_N[\v{x}]\;\; \v{\e}\,(\v{x})\q.\lb{8a}
\ee
In the basis $(\v{u}_s,\hat{e}_{\perp},\hat{e}_z)$, $M_N[\v{x}]$ is the
$3 \!\times \!3$ differential operator
\be
M_N[\v{x}] \;:=\; \L( \ba{ccc}
  -\Delta & 0 & 0 \\
0& -\Delta  \,+\, \L(\v{\nabla}\Psi_N\R)^2 &
 - 2\L(\v{\nabla}\Psi_N\R)\v{\nabla} \\[2mm]
0&    2\,\L(\v{\nabla}\Psi_N\R)\v{\nabla}    &
  -2 \Delta  \ea \R)\q.\lb{9}
\ee
The first component of the fluctuations generates small local rotations around
the axis $\v{u}_s$.
It corresponds to the spinwaves of the $SO(2)$ model and decouples from the
vortices as there.
The other two components couple to the vortices.
Without this coupling, the vortex part of the partition function
would be identical to the vortex part which one gets for the planar
XY model.
A priori, it is not clear, how the fluctuations influence the thermodynamics
of the vortices.
In order to study this second problem posed in the introduction, we need to
derive an effective description of the thermodynamics of the vortices.
To this end, we have to integrate the fluctuations in eqn.~\rf{8} in the
presence of the vortices, i.~e.~we have to calculate the functional
determinant of the semi positive elliptic operator $M_N[\v{x}]$.
This is a formidable problem, and to our knowledge, such a problem
has been solved only in a few cases. For example, Fateev
et.~al.~calculated
such functional determinants in their investigation of the thermodynamics of
the instantons in the $SO(3)$ nl$\s$ model \cite{ffs}.

Now, we first concentrate on the case of two vortices, $N=2$.
The contribution of the fluctuations to the effective interaction potential
of these two vortices be $\ln W := \ln \{\mbox{det}M_2/\mbox{det}M_0\}$.
In order to determine that part of the functional determinants which diverges
with the number of degrees of freedom, we used the proper-time regularization
and calculated the first nontrivial Seeley coefficients via a Heat-Kernel
expansion \cite{twoDQFT}.
The leading divergence cancels in numerator and denominator of $W$.
The next to leading (logarithmic) divergence vanishes exactly, i.~e.~$W$ is
finite.

We further evaluated $W$, using two different regularizations.
In the first, the analytical calculation, we observe that in bipolar
coordinates, the differential operators in $M_2$ separate.
The bipolar coordinates are $\ph$ and $\rho$ \cite{mf}, where $\ph(\v{x})
\equiv \Psi_2(\v{x})$ and $\rho(\v{x})$ is the orthogonal coordinate.
The x-y plane maps onto a cylinder of length
$2\ln\frac{r}{a}$ where $r=|\v{x}_1-\v{x}_2|$.
Furthermore, the coefficients of the differential operators in $M_2$ become
constant.
Gaussian integration of the eigenmodes of nonzero energy and exact integration
of the Goldstone modes over the group space of $SO(3)\!\times \!SO(2)$ yields
an expression for $\ln W$ containing sums over logarithms of the eigenvalues.
Extracting the asymptotic for $r \gg a$ we finally obtain
\be
\ln W^{asymp}_{bipol} \approx w \ln \frac{r}{a}
\,+\, \frac{1}{2}\ln \ln \frac{r}{a} \,+\,{\cal O}(1)\q.
\lb{10}
\ee
The expression for $w$ evaluates to 0.06.
This result shows that, within the harmonic approximation \rf{8a},
the coupling of the fluctuations to the vortices does not change the
asymptotic logarithmic behaviour of the interaction of the vortices.

We confirmed this result in a second regularization by calculating $\ln W$
numerically on finite lattices in real space.
We computed $\ln W$ for lattices up to sizes of $64\! \times \!64$.
For the values of $r$ ($r<64$) available, the results converge towards
the corresponding result of the analytic diagonalisation.
But, owing to the small value of $w$, one cannot reliably identify the
asymptotic $\ln \frac{r}{a}$ behaviour in the numerical data.
To obtain further confirmation of the agreement between the analytic and the
numerical calculation, we generalized the action \rf{0} by adding a term
$\a(\v{\nabla}\v{n}_3 )^2$ which does not change the symmetry (see \cite{awe}).
The calculations were then also performed for several values of $\a$ with
$0\leq \a<1$, where $\a=1$ is a system of the higher symmetry
$SO(3)\!\times \!SO(3)$.
Increase of $\a$ from 0 to 1 increases the prefactor $w$ in
\rf{10} smoothly to values of the order of unity.
This makes it possible to identify for $\a>0$ the $\ln \frac{r}{a}$ behaviour
in the numerical data even for small values of $r$. Especially the dependence
of $w$ on $\a$ shows a very good agreement between the analytic and the
numerical calculations.
However, the magnitude of the prefactor of the leading term in \rf{10} is less
important. The essential point is that, {\it within the harmonic
approximation, the fluctuations do not change the logarithmic form of the
interaction potential of two vortices at large distances}.

Finally, we turn to the case of $N>2$ vortices.
As in the $SO(2)$ model, the two--vortex excitations are the elementary
local excitations.
At large distances, the interaction between such excitations decays
as the inverse square of their distance and should not be changed by the
fluctuations.
With this assumption, we get for the general N--vortex case
\be
\ln \frac{ \cZ_{N}\L(\v{x}_1,\cdots,\v{x}_{N} \R)}{\cZ_0} = \frac{2\pi}{t'}
\sum\limits_{j \neq j'}\; s_j s_{j'} \;
               \ln \frac{|\v{x}_j-\v{x}_{j'}|}{a}\q.\lb{11}
\ee
\rf{11} together with \rf{8} leads to an expression for the partition function
which describes a 2-dimensional Coulomb gas \cite{kt}.
The connection between the dimensionless temperature $t$
and the effective temperature of the Coulomb gas $t'$ is:
$\frac{1}{t'}\;=\;\frac{1}{4t}\;( 1\;-\;\frac{w}{4\pi}t )$.\\

The representation \rf{8} with \rf{11} for the thermodynamics of the vortices
constitutes the result of our work. In the following summary, we want to
describe the implications for the classical HAFT.
For small $t$, the Coulomb gas \rf{11} is in its low temperature phase.
The Kosterlitz--Thouless transition temperature is $t'_{KT} \cong 2.7$.
Thus, one would expect a transition in the HAFT at around
$t_{KT} \cong 0.68$ (or $T_{KT}\cong 0.58 J S^2$).
However, in the steps leading to \rf{11}, anharmonic spinwave interaction
have been neglected, i.~e.~we considered a system of vortices in an ideally
correlated background.
As we have pointed out above, these anharmonicities lead to a finite
correlation length $\xi_{SW}$ even in the absence of vortices.
Hence, the validity of our results is restricted to distances smaller
than $\xi_{SW}$.
This, of course, implies that at $T_{KT}$ the system does not become critical.
Nevertheless, assuming that $\xi_{SW}$ is sufficiently
large, we expect a spontaneous creation and dissociation of pairs of
vortices to occur as the temperature increases towards $T_{KT}$, resulting in
a sudden breakdown of the remaining correlations.
The temperature dependence of the correlation length should then cross over
from the low temperature behaviour
$\xi_{SW} \sim \sqrt{t} \exp (1+{\ts \frac{\pi}{2}})\pi /t$
\cite{aza1} to the Kosterlitz--Thouless behaviour
$\xi_{KT} \sim \exp  b/{\ts\sqrt{t-t_{KT}}}$.
Regarding the difference in the value of our estimate of $T_{KT}$ and that
of MC simulations of Kawamura and Miyashita (\cite{km}:
$ T_{KM}\cong 0.31 \pm 0.01 J S^2$)
we refer to the similar situation in the case of the planar XY model.
Tobochnik and Chester found in their MC simulations of the planar XY model
\cite{tc} a value for the transition temperature $T_{KT}$ which is about
$30\%$ reduced from the value which was found from Kosterlitz and Thouless.

It is interesting to note that at $t_{KT}$ the spinwave correlation length
$\xi_{SW}$ would be very large ($\cong 10^5$ lattice spacings)
if one assumes the unknown prefactor to be of the order of unity as for the
classical HAFSQ \cite{hn}.
In recent MC calculations of the classical HAFT, Southern and Young
\cite{sy} found a sharp drop of the correlation length at a temperature
which agrees with the estimate $T_{KM}$ of Kawamura and Miyashita.
For $T>T_{KM}$, the data in \cite{sy} can be fitted well using $\xi_{KT}$.
The interpretation of this behaviour in ref.~\cite{sy} is the same as presented
here: there is a crossover in the correlation length of the HAFT
due to the unbinding of the vortices.\\

Acknowledgements: This work has been supported by the Deutsche
Forschungsgemeinschaft under grant no.~Ev 6/3-1.
The numerical calculations were carried out in part at the Regionales
Rechenzentrum Niedersachsen, Hannover.


\end{document}